\documentclass[twocolumn,aps,psfig]{revtex4}

\usepackage{graphicx}
\usepackage{dcolumn}
\usepackage{amsmath}

\begin{document}

\title{Homoclinic Chaos in the Dynamics of a General Bianchi IX Model}

\author{H. P. de Oliveira}
 \email{henrique@fnal.gov}
\affiliation{{NASA/Fermilab Astrophysics Center \\
Fermi National Accelerator Laboratory, Batavia, Illinois, 60510-500.}\\
and \\
{\it Universidade do Estado do Rio de Janeiro }\\
{\it Instituto de F\'{\i}sica - Departamento de F\'{\i}sica Te\'orica,}\\
{\it CEP 20550-013 Rio de Janeiro, RJ, Brazil}}

\author{A. M. Ozorio de Almeida}
\email{ozorio@cbpf.br}
\affiliation{{\it Centro Brasileiro de Pesquisas F\'\i sicas\\ Rua Dr.
Xavier Sigaud, 150 \\ CEP 22290-180, Rio de Janeiro~--~RJ, Brazil}\\
and \\
{\it Max Planck Institute for Physics of Complex Systems}\\
{\it Noethnitzer Strasse 38}\\
{\it 01187, Dresden, Germany.}}
\author{I. Dami\~ao Soares}
 \email{ivano@cbpf.br}
\affiliation{{\it Centro Brasileiro de Pesquisas F\'\i sicas\\ Rua Dr.
Xavier Sigaud, 150 \\ CEP 22290-180, Rio de Janeiro~--~RJ, Brazil}}
 \author{E. V. Tonini}
\affiliation{{\it Centro Federal de Educa\c c\~ao Tecnol\'ogica -
CEFETES Av.  Vit\'oria, 1729 -
Jucutuquara. CEP 29040-333, Vit\'oria~--~ES, Brazil} \\
and \\
{\it Centro Brasileiro de Pesquisas F\'\i sicas\\ Rua Dr.
Xavier Sigaud, 150 \\ CEP 22290-180, Rio de Janeiro~--~RJ, Brazil}}
\email{tonini@etfes.br}

\date{\today}










%



\begin{abstract}
The dynamics of a general Bianchi IX model with three scale factors is
examined. The matter content of the model is assumed to be comoving dust plus a
positive cosmological constant. The model presents a critical point of
saddle-center-center type in the finite region of phase space. This critical
point engenders in the phase space dynamics the topology of stable and unstable
four dimensional tubes $R \times S^3$, where $R$ is a saddle direction and
$S^3$ is the  manifold of unstable periodic orbits in the center-center sector.
A general characteristic of the dynamical flow is an oscillatory mode about
orbits of an invariant plane of the dynamics which contains the critical point
and a Friedmann-Robertson-Walker (FRW) singularity. We show that a pair of
tubes (one stable, one unstable) emerging from the neighborhood of the critical
point towards the FRW singularity have homoclinic transversal crossings. The
homoclinic intersection manifold has topology $R \times S^2$ and is constituted
of homoclinic orbits which are bi-asymptotic to the $S^3$ center-center
manifold. This is an invariant signature of chaos in the model, and produces
chaotic sets in phase space. The model also presents an asymptotic DeSitter
attractor at infinity and initial conditions sets are shown to have fractal
basin boundaries connected to the escape into the DeSitter configuration
(escape into inflation), characterizing the critical point as a chaotic
scatterer.
\end{abstract}

\maketitle


The longtime debate on the chaotic dynamics of general Bianchi IX models
started with the work of Belinskii, Khalatnikov and Lifshitz (BKL) on the oscillatory
behaviour of such models in their approach to the singularity\cite{BKL}. They
showed that the approach to the singularity($t\rightarrow 0$) of a general Bianchi IX
cosmological solution is an oscillatory mode, consisting of an infinite
sequence of periods (called Kasner eras) during which two of the scale
functions oscillate and the third one decreases monotonically; on passing from
one era to another the monotonic behaviour is transfered to another of the
three scale functions. The length of each era is determined by a sequence of
numbers ${X_s, 0<X_s<1,  s=integer}$, each of which arises from the preceding
one by the transformation $X_{s+1}$ = fractional part of $1/X_s$. From the
properties of this map it is obtained that the behaviour of the model becomes
chaotic on approaching the singularity($t=0$) for arbitrary initial conditions
given at $t_0>0$. With the advent of powerful numerical resources the interest
in the behaviour of these models revived  but - as in the BKL work - the
procedure has been basically to obtain maps which approximate the dynamics of
the model described by Einstein's equations and which exhibit strong stochastic
properties\cite{barrow}. How well these discrete maps represent the full
nonlinear dynamics has been subject of much research, particularly by
Berger\cite{berger1,berger2} and Rugh\cite{rugh}. The interest in the
chaoticity of Bianchi IX models has been mainly focused on the Mixmaster case
(vacuum Bianchi IX model with three scale factors\cite{misner}), but the
chaotic dynamics of other Bianchi IX model universes has also been discussed in
the literature (cf. \cite{burd} and references therein, and \cite{cornish});
homoclinic chaos in axisymmetric Bianchi IX universes with matter and
cosmological constant has been treated in \cite{barguine}. The chaoticity in
the Mixmaster dynamics has been object of much dispute in the literature (cf.
the contributions to the Section \textit{Bianchi IX (Mixmaster) dynamics} in
\cite{burd1}, particularly \cite{berger2}). Latifi et al.\cite{latifi} and
Contopoulos et al.\cite{contopoulos} have shown the nonintegrability of the
Mixmaster model in the Painlev\'e sense, although the question of the generic
behaviour (chaotic or not) remained unsettled mainly due to the absence of an
invariant (or topological) characterization of chaos in the model (standard
chaotic indicators as Liapunov exponents being coordinate dependent and
therefore questionable\cite{matsas}\cite{burd1}). More recently Cornish and
Levin\cite{levin} proposed to quantify chaos in the Mixmaster universe by
calculating the dimensions of fractal basin boundaries in initial conditions
sets for the full dynamics, these boundaries being defined by the code
associated with one of the three outcomes on which one of the three axes is
collapsing most quickly, as established \textit{numerically}. Their result
received a recent critical review that nevertheless endorses it\cite{letelier}.
In the present paper we shall give an invariant characterization of chaos for
the Mixmaster model with a positive cosmological constant, but this criterion
does not work for the case of zero cosmological constant.

Our purpose in this paper is to examine the dynamics of a general (three scale
factors) Bianchi IX cosmological model with dust and cosmological constant and
establish the existence of chaos using the criterion of the homoclinic
transversal crossing of topological tubes that organizes the dynamical flow in
the model. We show that the dynamics of the model is highly complex and
chaotic, and that chaos has a definite homoclinic origin. The phase space of
the model is noncompact and the presence of the cosmological constant
determines two crucial facts in phase space: first, the existence of a critical
point of the type saddle-center-center; second, two critical points at infinity
corresponding to the DeSitter configuration, one acting as an ``attractor" to
the dynamics. With respect to the latter point, our system has the
characteristic of a chaotic scattering system with two absolute outcomes
consisting of (i) escape into inflation (the DeSitter attractor) or (ii)
recollapse to the singularity. The presence of the critical point of
saddle-center-center type is responsible for a wealthy and complex dynamics,
engendering in phase space topological structures which are homoclinic to a
manifold of periodic orbits, having the topology of the three sphere $S^3$. The
dynamics has a 2-dim invariant plane allowing to expand the dynamical equations
about it, in the same procedure used to examine  the dynamics about a periodic
orbit of the system. The mixmaster universe with a cosmological constant will
correspond to the limiting case of dust equal to zero but, as we will discuss,
the skeleton of its dynamics is framed in the dynamics of the general case.

The line element of the model has the form

\begin{eqnarray}
\label{metric}
ds^2 &=& dt^2 - ({\theta^1})^2 - ({\theta^2})^2 -({\theta^3})^2 ,
\end{eqnarray}

\noindent where $t$ is the cosmological time and $ {\theta^1}= M(t){\omega^1}$,
${\theta^2}=N(t){\omega^2}$ and  ${\theta^3}=R(t){\omega^3}$. Here the
${\omega^i}$ are Bianchi IX 1-forms satisfying
$d{\omega^i}={\epsilon^{ijk}}{\omega^j}\wedge{\omega^k}$. The matter content of
the models is assumed to be a pressureless perfect fluid, namely dust, with
energy density ${\rho}$, as described by the comoving observers with four
velocity orthogonal to the homogeneity surfaces of (\ref{metric}), plus a
positive cosmological constant $\Lambda$. The dynamics of the three scale
factors M(t), N(t) and R(t) will be given by Einstein's equations, which are
equivalent to Hamilton's equations generated by the Hamiltonian constraint
\begin{eqnarray} \label{hamilton} \nonumber H &=&
\frac{1}{8}\,\left(-\frac{M}{NR}p_M^2-\frac{N}{MR}p_N^2\right.\\ \nonumber
&-&\left.\frac{R}{MN}p_R^2+\frac{2}{R}p_M p_N +\frac{2}{M}p_N
p_R+\frac{2}{N}p_M p_R\right)\\ \nonumber &+ & \frac{1}{2MNR}(R^4+M^4+N^4
-(R^2-N^2)^2 -(R^2-M^2)^2\\ &-&(M^2-N^2)^2) - 2\Lambda MNR - 2 E_0 = 0.
\end{eqnarray} Here $p_M$, $p_N$ and $p_R$ are the momenta canonically
conjugate to $M$, $N$ and $R$, respectively. $E_0$ is a constant proportional
to the total energy of the model, arising from the first integral of the
Bianchi identities,  $\rho MNR=E_0$. Hamilton's equations have the form

\begin{eqnarray}
\label{eq3}
\nonumber
\dot{M} &=& \frac{\partial H}{\partial p_M}\,=
\left(-\frac{M}{4NR}p_M+\frac{1}{4R}p_N +\frac{1}{4N}p_R \right),\\
\end{eqnarray}

\begin{eqnarray}
\label{eq4}
\nonumber
\dot{p}_M &=& -\frac{\partial H}{\partial M}\,=\left(\frac{1}{8NR}p_M^2-\frac{N}{8M^2R}p_N^2\right.\\ \nonumber
&-&\left.\frac{R}{8M^2N}p_R^2+\frac{2}{8M^2}p_N p_R\right)\\ \nonumber
&+ & \frac{1}{2M^2NR}(R^4+M^4+N^4 -(R^2-N^2)^2\\
&-&(R^2-M^2)^2 -(M^2-N^2)^2) + 2\Lambda NR\\ \nonumber 
&-&\frac{1}{2MNR}(4M^3+4(R^2-M^2)M\\ \nonumber
&-&4 (M^2-N^2)M),
\end{eqnarray}

\noindent and cyclically in $M$, $N$, $R$, $p_M$,$p_N$, $p_R$. The dynamical
system (\ref{eq3}) and (\ref{eq4}) presents one critical point E with
coordinates

\begin{equation}
\label{eq5}
p_M = p_N = p_R = 0, \quad M = N = R = M_0, 
\end{equation}
where $M_0 = \sqrt{\frac{1}{4\Lambda}}$. The energy of the critical point E is defined by

\begin{eqnarray}
\label{eq6}
2 E_{crit} =\sqrt{\frac{1}{4\Lambda}}.\\ \nonumber
\end{eqnarray}

Much of our understanding of nonlinear systems derives from linearization about
critical points and the determination of existing invariant submanifolds, which
are structures that actually organize the dynamics in phase space. The system
presents a two dimensional invariant manifold of the dynamics,  defined by

\begin{equation}
p_M=p_N=p_R,\;\; M=N=R. \label{eq7}
\end{equation}

\noindent This invariant plane is actually the intersection of the two four
dimensional invariant submanifolds defined by ($M=N,p_M=p_N$) and ($N=R,
p_N=p_R$). The saddle-center-center critical point $E$ belongs to the invariant
plane, and from it emerge the separatrices $S$. The FRW singularity
($M=N=R=0$,$p_{M}=p_{N}=p_{R}=0$) is a degenerate critical point of the
dynamics on the invariant plane. The phase picture of the motion in the
invariant plane is given in Fig. 1, already expressed in canonical
coordinates $(x, p_x)$, to be introduced in Eqs. (\ref{canon1})-(\ref{canon2}),
that are variables defined on the invariant plane. Also, a straightforward
analysis of the infinity of the phase space under consideration shows that it
has two critical points in this region, corresponding to the DeSitter solution,
one acting as an attractor (stable DeSitter configuration) and the other as a
repeller (unstable deSitter configuration) for the dynamics at infinity. The
scale factors $M$, $N$ and $R$ approach the DeSitter attractor as $M=N=R\sim
exp[t\sqrt{\Lambda/3}]$ and $p_M=p_N=p_R\sim exp[2t\sqrt{\Lambda/3}]$. It is
easy to see that the DeSitter asymptotic configurations also belong to the
invariant plane and that two of the separatrices $S$ approach them for times
going to $\pm\infty$.

\begin{figure}
\rotatebox{270}{\includegraphics*[scale=0.3]{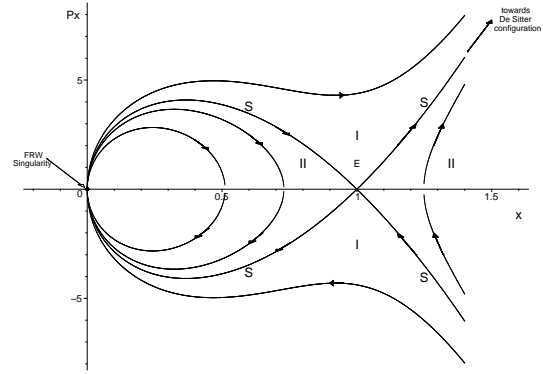}}
\vspace{0.5cm}
\caption{Phase portrait of the invariant plane $M=N=R, p_M=p_N=p_R$, in the
coordinates
$(x,p_x)$ defined by $(21)-(22)$. The separatrices S are characterized by the
energy $E_0=E_{crit}=\frac{1}{4\sqrt{\Lambda}}$. $E$ is the critical point of
saddle-center-center type, that belongs to the invariant plane
(cf. text). $x=0, p_x=0$ is the FRW singularity.}
\end{figure}

To proceed with the study of the phase space dynamics, let us now linearize
the dynamical equations (\ref{eq3}), (\ref{eq4}) about the critical point E. We
define

\begin{eqnarray}
\label{eq8}
\nonumber
X &=& (M - M_0),   W = (p_M - 0), \\
Y &=& (N - M_0),   K = (p_N - 0), \\ \nonumber
Z &=& (R - R_0),   L = (p_R - 0).\\ \nonumber 
\end{eqnarray}

\noindent We obtain

\vspace{-0.0cm}

\begin{equation}
\nonumber
\left(\begin{array}{c}
      \dot{X} \\
      \dot{Y} \\
      \dot{Z} \\
      \dot{W} \\
      \dot{K} \\
      \dot{L} \\
\end{array} \right) = \frac{1}{M_0}\ S \left(\begin{array}{c}
      X \\
      Y \\
      Z \\
      W \\
      K \\
      L \\
\end{array} \right)
\end{equation}


%
%
%
%
%

\noindent where

\begin{equation}
\nonumber
\left(\begin{array}{cccccc}
	0 & 0  & 0  & -1/4 &  1/4 &  1/4 \\
	0 & 0  & 0  &  1/4 & -1/4 &  1/4 \\
        0 & 0  & 0  &  1/4 & 1/4  & -1/4 \\
	3 & -1 & -1 & 0    & 0    & 0    \\
       -1 & 3  & -1 & 0    & 0    & 0    \\
       -1 & -1 & 3  & 0    & 0    & 0
\end{array} \right)
\end{equation}


\noindent In what follows, without loss of generality, we fix $\Lambda$ such
that $M_0=1$. The nature of the critical point E is determined by the
characteristic polynomial associated with the linearization matrix $S$. We
obtain

\begin{eqnarray}
\label{eq9}
P(\lambda) = ({\lambda}^2 -\frac{1}{4})({\lambda}^2 + 2)^{2},\\ \nonumber
\end{eqnarray}
with roots

\begin{eqnarray}
\label{eq10}
\lambda = \pm \frac{1}{2} ,      \lambda= \pm i\sqrt {2},\\
\nonumber
\end{eqnarray}
where the second pair has multiplicity two. The pair of real eigenvalues
generates a saddle structure, while the degenerate pair of imaginary
eigenvalues generate a double center structure. The analysis of the center
structure will reveal a manifold of linearized unstable periodic orbits with
the topology of a $3$-sphere, as we will see.

To display the structure of the linearized motion, we start by diagonalizing
$S$ with the use of a transformation matrix $\Re$ whose columns are composed of
six independent eigenvectors of $S$ \cite{siegel}. A judicious choice of  $\Re$
yields primed variables defined by the transformation

\begin{eqnarray}
\label{eq11}
\nonumber
{X}^{\prime}&=&\frac{1}{3}(X+Y+Z),\\ \nonumber
{Y}^{\prime}&=&(X-Y),\\ 
{Z}^{\prime}&=&(X+Y-2Z),\\ \nonumber
{W}^{\prime}&=&(W+K+L),\\ \nonumber
{K}^{\prime}&=&\frac{1}{2}(W-K),\\ \nonumber
{L}^{\prime}&=&\frac{1}{6}(W+K-2L).\\ \nonumber
\end{eqnarray}

\noindent In these new variables, the quadratic Hamiltonian about $E$ is
expressed in the form

\begin{eqnarray}
\label{eq12}
H = (E_{crit} - E_0) +\frac{1}{4}
\left(\frac{1}{6} W^{\prime 2} - 6 X^{\prime 2}\right)\\
-\left(\frac{1}{2}K^{\prime 2}+Y^{\prime 2}\right) -
\left(\frac{3}{2} L^{\prime 2} + \frac{1}{3} Z^{\prime 2}\right).
\nonumber
\end{eqnarray}

\noindent These variables are conjugated to the pairs according
to $[X^{\prime},W^{\prime}]=1, [K^{\prime},Y^{\prime}]=1,
[L^{\prime},Z^{\prime}]=1$, other Poisson brackets (PB) zero. The
Hamiltonian (\ref{eq12}) is separable, and we can immediately identify the
following constants for the linearized motion

\begin{eqnarray}
E_{hyp}=\frac{1}{4}\left(\frac{1}{6} W^{\prime 2} - 6 X^{\prime 2}\right),\\
E_{rot_{1}}=\frac{3}{2} L ^{\prime 2}+\frac{1}{3} Z^{\prime 2},\\
E_{rot_{2}}=\frac{1}{2} K^{\prime 2} + Y^{\prime 2},\\
Q_1=\frac{1}{3}Y^{\prime} Z^{\prime} + \frac{1}{2} K^{\prime} L^{\prime},\\
Q_2=L^{\prime} Y^{\prime} - \frac{1}{3} Z^{\prime} K^{\prime},\\
\nonumber
\end{eqnarray}

\noindent in the sense that they all have zero PB with the Hamiltonian
(\ref{eq12}). The first three constants appear already as separable pieces in
the Hamiltonian (\ref{eq12}). $E_{hyp}$ corresponds to the energy associated
with motion in the saddle sector, while $E_{rot_{1}}$ and $E_{rot_{2}}$ are the
rotational energies associated with the motion in the linear part of the
center-center manifold, namely, on stable periodic orbits in 2-dim tori. The
remaining two constants are additional symmetries that arise as consequence of
the multiplicity two of the imaginary eigenvalues, and are associated to the
fact that the linearized dynamics in the center-center sector is that of a
two-dimensional isotropic harmonic oscillator. They are not all independent but
are related by

\begin{eqnarray} \label{eq18}
4E_{rot_{1}} E_{rot_{2}} = 12 Q_1^2 + 6 Q_2^2.
\end{eqnarray}

We are now ready to describe the topology of the general dynamics in a linear
neighborhood of the critical point $E$. To describe all possible motions, the
following situations must be taken into account in the linearized Hamiltonian
(\ref{eq12}). If $E_{hyp}=0$ two possibilities arise.  First, we have
${W}^{\prime}={X}^{\prime}=0$, implying that the motions are periodic orbits of
the two-dimensional isotropic harmonic oscillator

\begin{eqnarray}
\label{eq19}
H =\left(\frac{1}{2} K^{\prime 2}+ Y^{\prime 2}\right) +
\left(\frac{3}{2} L^{\prime 2} + \frac{1}{3} Z^{\prime 2}\right)\\ \nonumber
=(E_{crit} - E_0).
\end{eqnarray}

\noindent By a \textit{proper canonical reescaling} of the variables in
(\ref{eq11}) and (\ref{eq19}) it is easy to see that these constant energy
surfaces are hyperspheres and that the constants of motion $Q_1$, $Q_2$ and
$Q_3=(E_{rot_{1}}-E_{rot_{2}})$ satisfy the algebra of the three dimensional rotation
group under the Poisson bracket operation, namely,

\begin{eqnarray}
\label{eq20}
[Q_i,Q_j]=\epsilon^{ijk}Q_k
\end{eqnarray}

\noindent The constant of motion $Q_1$ considered as a generator of
infinitesimal contact transformations has a peculiar significance in
characterizing the topology of the underlying group of the algebra
(\ref{eq20})\cite{mcintosh}. While $Q_2$ generates infinitesimal rotations of
the orbits, $Q_1$ generates infinitesimal changes in eccentricity. The action
of $Q_1$ is to take an orbit - let us say nearly circular - and transforms it
into an orbit of higher and higher eccentricity until it collapses into a
straight line. Continued application of $Q_1$ produces again an elliptic orbit,
but now traversed in the opposite sense , so that it takes a $720^{0}$ rotation
to bring the orbit back into itself. The two-valuedness of the mapping arises
from the fact that the orbits are oriented. Therefore we are led to the
conclusion that the group generated by these constants of motion is the unitary
unimodular group and not the rotation group. It is therefore compatible to
consider that the center-center manifold has indeed the topology of a three
sphere $S^{3}$. Now the separate conservation of $E_{rot_{1}}$ and $E_{rot_{2}}$
(cf.Eq.(\ref{eq19})) allows us to show that the center-center manifold in the
linear neighborhood of $E$ is foliated by Clifford two-dimensional surfaces in
$S^{3}$ \cite{clifford}, namely,  two-tori $\Im_{E_0}$ contained in the energy
surface $E_0=const$. Such surfaces, as well as the $S^3$ manifold that contains
them depend continuously on the parameter $E_0$. We remark that these two tori
will have limiting configurations which are periodic orbits, whenever
$E_{rot_{1}}=0$ or $E_{rot_{2}}=0$, and correspond to the case of maximum
eccentricity (for instance, a straight line in the plane
$({Y}^{\prime},{Z}^{\prime}))$.

The second possibility will be ${W}^{\prime}=\pm 6 {X}^{\prime}$, that defines
the linear stable $V_S$ and unstable $V_U$ manifolds of the saddle-sector.
$V_S$ and $V_U$ limit regions $I$ ($E_{hyp}<0$) and regions $II$ ($E_{hyp}>0$)
of motion on hyperbolae which are solutions in the separable saddle-sector (cf.
(\ref{eq12}))
$E_{hyp}=\frac{1}{4}\left(\frac{1}{6} W^{\prime2}-6 X^{\prime2}\right)$, (cf.
(\ref{eq12})). Note that the saddle-sector depicts the structure of Fig. 1 in
the neighborhood of $E$, with $V_U$ and $V_S$ tangent to the separatrices at
$E$ \cite{barguine}. The direct product of $\Im_{E_0}$ with $V_S$ and $V_U$
generates, in the linear neighborhood of $E$, the structure of stable
$(\Im_{E_0}\times V_S)$ and unstable $(\Im_{E_0}\times V_U)$ 3-dim tubes, which
coalesce into the two-dimensional tori $\Im_{E_0}$ for times going to $+\infty$
or $-\infty$, respectively. The energy of any orbit on these tubes is the same
as that of the orbits on the two tori $\Im_{E_0}$. These 3-dim tubes are
confined inside the 4-dim tubes which are the product of $V_S$ and $V_U$ by the
$S^3$ center-center manifold. It is obvious that the existence of the $S^3$
center-center manifold is restricted to energy surfaces such that $(E_{crit} -
E_0)>0$. The intersection of the center-center manifold with the energy surface
$H=E_0$ is in general a three sphere parametrized with $E_0$, from which two
pairs (one stable and one unstable) of 4-dim tubes emanate (cf. Fig. 2 and
caption). More particularly, in the linear neighborhood of $E$ (namely, for
$(E_{crit}-E_0)$ small), two pairs of 3-dim tubes emanate from the Clifford
two-dimensional surface $\Im_{E_0}$ contained in $S^{3}$. For the case
$E_{hyp}\not=0$, the motion is restricted on infinite tubes resulting from the
direct product of the hyperbolae, lying in the regions $I$ and $II$ of Fig. 1,
with $\Im_{E_0}$. A general orbit which visits the neighborhood of $E$ belongs
to the general case $E_{hyp}\not=0$, $E_{rot_{1}}\not=0$ and $E_{rot_{2}}\not=0$. In
this region the orbits have an oscillatory approach to the 3-dim linear tubes,
the closer as $E_{hyp}\rightarrow0$.

\begin{figure}
\rotatebox{270}{\includegraphics*[scale=0.4]{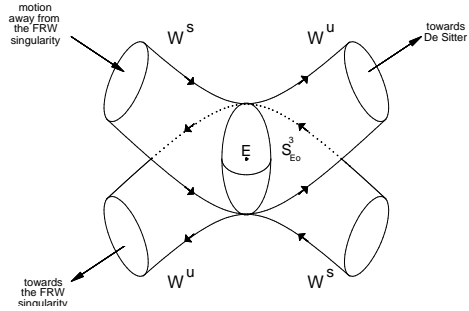}}
\\
\vspace{0.5cm}
\caption{Stable and unstable 4-dimensional tubes emanating
from the center-center manifold $S^3_{E_0}$. They are the nonlinear extension
of $S^3_{E_0}(\Im_{E_0})\times V_S$ and $S^3_{E_0}(\Im_{E_0})\times V_U$ in the
neighborhood of $E$.}
\end{figure}

We remark that the surface of the tubes constitutes a boundary for the general
flow and, in the neighborhood of $E$, is defined  by $E_{hyp}=0$. Depending on
the sign of $E_{hyp}$ the motion will be confined inside the 4-dim tube (for
$E_{hyp}<0$) and will correspond to a flow of motion distinct from the one
confined outside the tube (for $E_{hyp}>0$). This can be seen from
Eq.(\ref{eq12}), that can be rewritten \begin{eqnarray} \nonumber
E_{rot_{1}}+E_{rot_{2}}-E_{hyp}=(E_{crit}-E_{0}). \end{eqnarray}

\noindent Let us consider, for instance, a set of initial conditions
corresponding to initially expanding universes. Two distinct flows will be
associated with these initial conditions, depending whether they are contained
inside the tube or outside the tube. A careful examination of Fig. 2 shows us
that the flow corresponding to initial conditions inside the stable tube will
reach a neighborhood of $E$ and will return towards the FRW singularity inside
the unstable tube (with $E_{hyp}<0$), while the flow of orbits associated to
initial conditions that are outside the tube will reach the neighborhood of $E$
and escapes towards the DeSitter attractor along the exterior of the unstable
tube that is directioned to the right. Also the flow of the stable tube on the
bottom right of Fig. 2 will escape towards the DeSitter attractor along the
interior of the unstable tube of the top right. We remark that the direction of
the separatrices and hyperbolae in the linearized saddle-sector are the
topological guide for the flow. Our main interest will rely on the stable and
unstable tubes that emerge from the neighborhood of the critical point $E$
towards the FRW singularity and their homoclinic transversal crossings, that
constitutes an invariant signature of chaos in the models.

By continuity, the nonlinear extension of the center-center
manifold will maintain the topology of the three-sphere $S^3$, but will not be
decomposable into $E_{rot_{1}}$ and $E_{rot_{2}}$ so that now only the four
dimensional tubes with topology $R \times S^{3}$ are meaningful for the
nonlinear dynamics.

The extension of the structure of the four dimensional tubes away from the
neighborhood of $E$ are now to be examined, and our basic interest will reside
in on the stable and unstable pair, $S^3_{E_0}\times V_S$ and $S^3_{E_0}\times
V_U$, that leave the neighborhood of $E$ and proceed towards the
Friedmann-Robertson-Walker (FRW) singularity of the model. To this end let us
consider the two dimensional invariant manifold of the dynamics, defined by
Eqs.(\ref{eq5}). The two dimensional invariant plane is contained in a six
dimensional phase space and it is obvious that, contrary to examples in lower
dimensional systems, it does not separate the phase space in two disjoint
parts. In fact the general motion about a neighborhood of the invariant plane
is an oscillatory flow confined along the interior or the exterior of the four
dimensional tubes $R\times S^3$, where the invariant plane (or more properly,
one of the curves of the invariant plane) may be thought as a structure in the
center of the tube, as we proceed to show. To see this, let us introduce the
canonical coordinate transformation with the generating function

\begin{eqnarray}
\nonumber
G=(MNR)^{\frac{1}{3}} p_x +\frac{M}{N} p_y + \frac{MN}{R^{2}} p_z \\
\nonumber
\end{eqnarray}


\noindent where $p_x$, $p_y$ and $p_z$ are the new momenta, resulting in

\begin{eqnarray}
x=(MNR)^{\frac{1}{3}}, y=\frac{M}{N}, z=\frac{MN}{R^{2}} \label{canon1}
\end{eqnarray}


\noindent and

\begin{eqnarray}
\label{canon2}
\nonumber
p_M=\frac{1}{3}\frac{NR}{(MNR)^{\frac{2}{3}}}p_x + \frac{1}{N}p_y +\frac{N}{R^{2}}p_z\\
p_N=\frac{1}{3}\frac{NR}{(MNR)^{\frac{2}{3}}}p_x - \frac{M}{N^{2}}p_y +\frac{M}{R^{2}}p_z\\
\nonumber
p_R=\frac{1}{3}\frac{NR}{(MNR)^{\frac{2}{3}}}p_x - \frac{2MN}{R^{3}}p_z.\\
\nonumber
\end{eqnarray}

\noindent It is worth remarking that the linearization of this canonical
transformation about the critical point $E$ yields exactly the linear
transformation (\ref{eq11}), and that the variables $(y,p_y,z,p_z)$  correspond
to the primed variables
$({K}^{\prime},{Y}^{\prime},{L}^{\prime},{Z}^{\prime})$ defined on the $S^3$
center-center manifold about a linear neighborhood of the critical point. The
variable $x$ is obviously the average scale factor of the model. In these new
variables, the full Hamiltonian (\ref{hamilton}) assumes the form

\begin{eqnarray}
\label{eq23}
\nonumber
H=\frac{1}{24x}p_x^{2}-\frac{y^{2}}{2x^3}p_y^{2}-\frac{3z^{2}}{2x^3}p_z^{2}\\
\nonumber
-\frac{x}{2z^{\frac{4}{3}}}-\frac{1}{2}x z^{\frac{2}{3}}y^{2}-\frac{1}{2y^{2}}x z^{\frac{2}{3}}\\
+\frac{x}{yz^{\frac{1}{3}}}+\frac{yx}{z^{\frac{1}{3}}}+x z^{\frac{2}{3}}
-2{\Lambda}x^{3}=2E_0,\\
\nonumber
\end{eqnarray}

\noindent and the equations of the invariant plane reduce to

\begin{equation}
\label{eq24}
y=1,z=1,\quad p_y=0=p_z.
\end{equation}

\noindent It is clear that $(x,p_x)$ are variables defined on the invariant
plane, and now the expansion of the Hamiltonian (\ref{eq23}) about the
neighborhood of the invariant plane can be easily implemented, producing a
linearized Hamiltonian parametrized by the variables $(x(t),p_{x}(t))$
describing the curves in the invariant plane, in analogy with the way we expand
a dynamical system about a periodic orbit. We obtain

\begin{eqnarray}
\label{eq25}
\nonumber
H=\frac{1}{24x}p_x^{2}+\frac{3x}{2}-2{\Lambda}x^3-\frac{1}{2x^3}p_y^{2}-\frac{3}{2x^3}p_z^{2}\\
-x(y-1)^{2}-\frac{1}{3}x(z-1)^{2}=2E_0
\end{eqnarray}

\noindent with dynamical equations

\begin{eqnarray} \label{eq26}
\nonumber
(\delta y \dot{)}=-\frac{1}{x^{3}}{\delta}p_y\\
\nonumber
(\delta p_y \dot{)}=2x{\delta}y\\
(\delta z \dot{)}=-\frac{3}{x^{3}}{\delta}p_z\\
\nonumber
(\delta p_z \dot{)}=\frac{2x}{3}{\delta}z,\\
\nonumber
\end{eqnarray}


\noindent where ${\delta}y=y-1$, ${\delta}z=z-1$, ${\delta}p_y=p_y-0$ and
${\delta}p_z=p_z-0$. The linearization matrix in (\ref{eq26}) has imaginary
eigenvalues, both with multiplicity two, given by $\lambda= \pm \frac{i\sqrt
{2}}{x(t)}$, so that in the neighborhood of the invariant plane we have only
elliptic modes, namely, the motion is oscillatory about the invariant plane
with increasing frequency as the orbits approach $x=0$. Close to the critical
point $E$, the motion in the invariant plane is slow, i.e.,
$(\dot{x},\dot{p_x})$ is small. Following an orbit in this plane with
diminishing $x$, its phase space speed increases but not faster than the
frequency of the surrounding motion in the other four variables. We are thus
justified in deriving a qualitative picture for the motion near the invariant
plane from an adiabatic approximation for the relevant orbit manifolds. That
is, we will assume that the overall geometry of the four dimensional stable and
unstable tubes that are asymptotic to the $S^3$ center-center manifold can be
approximated by the adiabatic deformation of this sphere as $x$ decreases away
from the critical point. In this way, we have in the six dimensional phase
space the four dimensional tubes of motion $R\times{S^3}$ with the two
dimensional invariant plane as the structure at its center, about which are the
oscillatory degrees of freedom $(y, p_y, z, p_z)$. We will use this fact to
show that the motion in the invariant plane guides the four dimensional tubes
of motion inducing inevitably the crossing of the unstable tube with the stable
one in the neighborhood of $(x=0, p_x=0)$. Also the eigenvectors of the matrix
in (\ref{eq26}) can give us an idea of the behaviour of the flow in the tubes
as $x$ diminishes, although in the neighborhood of $x=0$ the linear expansion
(\ref{eq25})-(\ref{eq26}) is no longer valid. Associated with the eigenvalue
$\lambda= \frac{i\sqrt {2}}{x}$ we may choose the two independent eigenvectors

\begin{eqnarray}
\nonumber
e_1&=&\left[\frac{1}{x},\frac{2x}{i\sqrt{2}}, 0, 0\right]\\
\nonumber
e_2&=&\left[0, 0, \frac{1}{x}, \frac{2x}{3i\sqrt{2}}\right],\\
\nonumber
\end{eqnarray}


\noindent and for the eigenvalue $\lambda= -\frac{i\sqrt {2}}{x}$,

\begin{eqnarray}
\nonumber
e_3&=&\left[-\frac{i\sqrt{2}}{2x}, x, 0, 0\right]\\
\nonumber
e_4&=&\left[0, 0, -\frac{3i\sqrt{2}}{2x}, x\right].\\
\nonumber
\end{eqnarray}


\noindent From the above form of the eigenvectors, we can draw a series of
important informations about the behaviour of the four dimensional tubes as
they approach the singularity $x=0$. As $x$ decreases, the sections $x=const.$
of the four dimensional tubes about the invariant plane stretches in two
directions and contracts in other two, while maintaining their symplectic
invariants constant. Also as $x$ decreases the oscillations have increasing
instantaneous frequency $\propto\frac{\sqrt{2}}{x}$. This appears as a
characteristic of the dynamical approach to the singularity even for the
extension of the tubes to a nonlinear neighborhood of the invariant plane.
There is a further fundamental fact that results from the $x$-dependence of the
above eigenvectors: let us consider for instance the sections $x=1$ of the
tubes which emanate from the two-tori $\Im_{E_0}$ defined in the linear
neighborhood of the critical point $E$. Their image projected on the planes
$(y, p_y)$, $(z, p_z)$ can be rescaled to small circles that deforms into
ellipses, as $x$ is infinitesimally diminished; the ellipse is rotated
infinitesimally counterclockwise if the section corresponds to the inferior
branch of the curves in the invariant plane ($p_x<0$) or clockwise if it
corresponds to the superior branch ($p_x>0$). This will be fundamental to
garantee the \textit{transversal} crossing of the tubes in the neighborhood of
$x=0$, as we will discuss. We must however remark that this conclusion is taken
in the adiabatic approximation, namely, considering that the variation of $x$
with time is adiabatic; this certainly does not occur near the singularity but
we assume that no drastic change in the dynamics will alter this behaviour for
the actual dynamical tubes. We illustrate this numerically in Fig. 3 by
depicting the sections $x=0.5$ projected on the planes $(y,p_y)$.
Figs. 4 show a numerical illustration of the unstable and stable 4-dim tubes
emanating from the neighborhood of $E$ towards the FRW singularity. We note
that, for $E_{crit}-E_0$ small, the motion proceeds towards the singularity
along a tube whose projection on the invariant plane $(x,p_x)$ \textit{shadows}
the invariant curve corresponding to $H=E_0$, along the upper branch and the
lower branch (cf.Fig. 4(b)).

We proceed now to show the homoclinic crossing of the stable and unstable
tubes at the neighborhood of the singularity $(x=0, p_x=0)$, a phenomenon
analogous to Poincar\'e's homoclinic tangle, and source of chaos in the
model\cite{koiller}. The four following points are essential for this. First,
in the five dimensional energy surface $H=E_0$, the four dimensional tubes of
motion $R \times S^{3}$ actually divide the space, separating the dynamics
outside from the dynamics inside the tubes. Second, in the canonical variables
$(x, p_x, y, p_y, z, p_z)$ introduced above, we have seen that $(x, p_x)$ are
the variables on the invariant plane, whose phase space picture is depicted in
Fig. 1, and the remaining degrees of freedom $(y, p_y, z, p_z)$ correspond to
elliptic modes of motion (cf. analysis in the neighborhood of the invariant
plane). Actually, the tubes have the two dimensional invariant plane as the
structure at its center (more properly, one of the curves of the invariant
plane corresponding to $H=E_0$), about which the flow with the oscillatory
degrees of freedom $(y, p_y, z, p_z)$ proceeds and in fact the motion in the
invariant plane \textit{guides} the flow, inducing inevitably the crossing of
one of the four-dimensional unstable tube with the stable one in a neighborhood
of $(x=0, p_x=0)$. Third, we note that the invariant plane cannot leave the
interior of the four dimensional tubes since they are separation surfaces in
the five dimensional manifold $H=E_0$, and fourth the property of clockwise
rotation or counterclockwise rotation of the sections of the stable and
unstable tubes, respectively, in their approach to the FRW singularity. Summing
up, in the five dimensional energy surface $H=E_0$, the model presents the
structure of four dimensional tubes of orbits, emanating from a three sphere
along the two saddle directions about the critical point saddle-center-center
$E$, one stable and one unstable (the stable one corresponding to initially
expanding universes). These tubes have the invariant plane curve as the
structure at its center and the motion in the invariant plane \textit{guides}
the oscillatory flow towards the singularity $(x=0, p_x=0)$, inducing
inevitably the crossing of the unstable tube with the stable one in a
neighborhood of the singularity; that the crossing is not just a point comes
from the conservation of sympletic areas in Hamiltonian dynamics, and the
transversality of the crossing comes from the clockwise or counterclockwise
rotation of the sections $(y,p_y)$ and $(z,p_z)$ of the stable and unstable
tubes, respectively. If we consider the transversal crossing in a section, say
$x=const.$, it is not difficult to see that the intersection is a $S^2$
manifold. Therefore the intersection manifold will be a 3-dim tube of flow
(with topology $R \times S^{2}$) which is contained \textit{both in the 4-dim
stable tube and in the 4-dim unstable tube}, and homoclinic to the $S^3$
center-center manifold. It is the equivalent of a homoclinic 1-dim orbit in
lower dimensional cases.

\begin{figure}[htb]
\rotatebox{270}{\includegraphics*[scale=0.3]{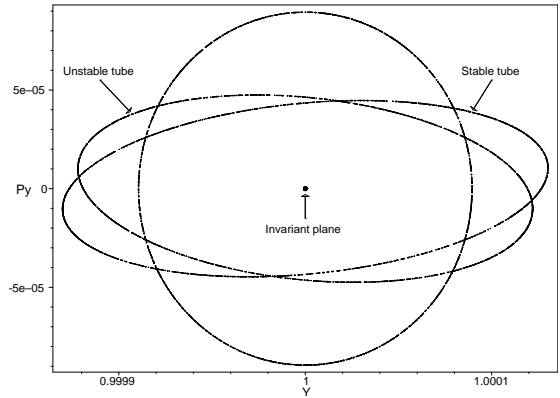}}
\vspace{0.5cm}
\caption{Sections of the stable and unstable tubes projected on the plane $(y,p_y)$.
The dotted circle is the projection of the 2-dim torus $\Im_{E_0}$ defined in
the linear neighborhood of $E$. The ellipses are the sections for $x=0.5$
of the tubes emanating from $\Im_{E_0}$. Note the distortion of the circle into
the ellipses and the clockwise/counterclockwise rotation of the sections of the
unstable/stable 3-dim tubes as $x$ diminishes (from $x=1$ to $x=0.5$) towards
the FRW singularity.}
\end{figure}

\begin{figure}[htb]
\rotatebox{270}{\includegraphics*[scale=0.3]{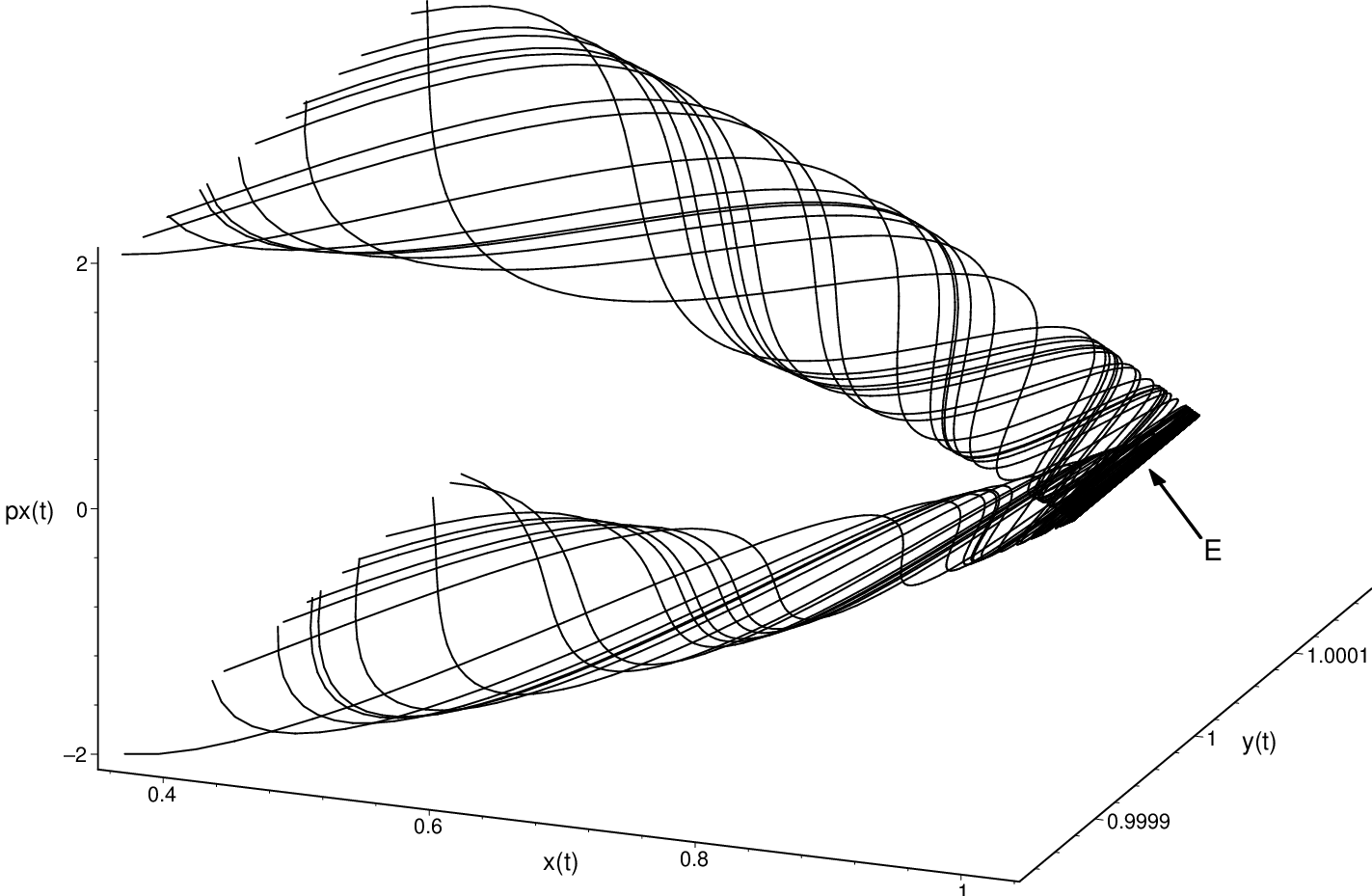}}
\vspace{0.7cm}
\centerline{\small{Fig. 4(a)}}
\vspace{0.7cm}
\rotatebox{270}{\includegraphics*[scale=0.3]{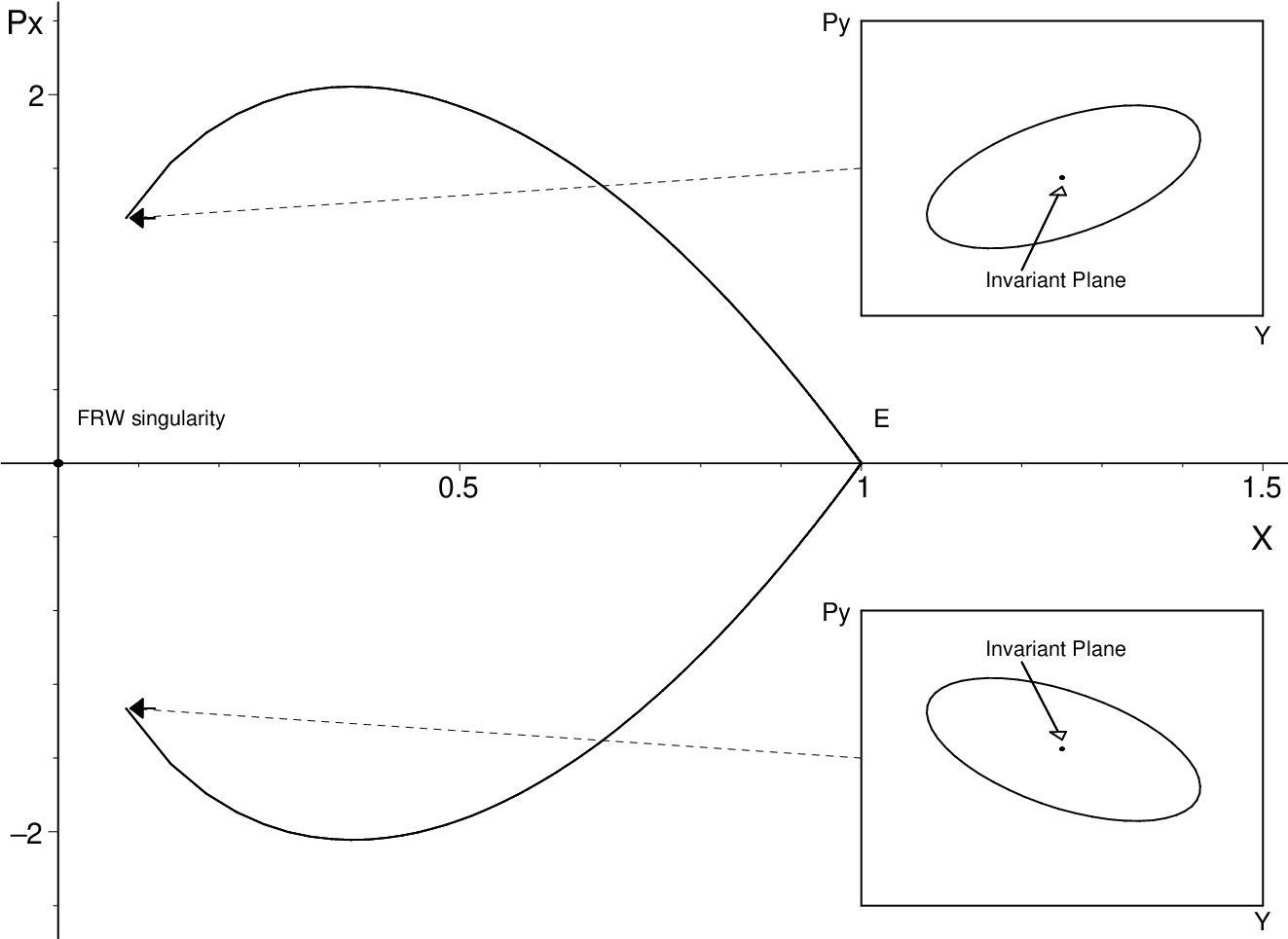}}
\centerline{\small{Fig. 4(b)}}
\vspace{0.5cm}
\caption{(a) Numerical illustration of the stable and unstable tubes emanating
from the neighborhood of the critical point $E$ towards the FRW singularity,
projected on the submanifold $(x,y,p_x)$, for $E_{crit}-E_0=10^{-8}$. (b)
Projection of the same stable and unstable tubes of (a) on the invariant plane
$(x,p_x)$. Note that the projection ``shadows" the
orbits on the invariant plane corresponding to $H=E_0$. The insets show the
sections $x=0.2$ projected  on the $(y,p_y)$ plane.}
\end{figure}

Typically, if the 4-dim tubes intersect transversally once they will intersect
each other an infinite number of times producing an infinite set of homoclinic
\textit{orbits} which are actually topological 3-dim tubes $R \times S^{2}$.
The homoclinic 3-dim tube, which is bi-asymptotic to the $S^3$ manifold
provides the mechanism for stretching and contraction, giving origin to the
homoclinic tangle, \textit{which is an invariant signature of chaos in the
model} \cite{wiggins}. We just mention that the dynamics near homoclinic
\textit{orbits} is very complex, forming well-known horseshoe structures (cf.
\cite{wiggins},\cite{koiller},\cite{ozorio} and \cite{ozorio1}, and the
bibliography therein).

Let us consider the first intersection of the tubes; a part of the orbits
inside the unstable tube will enter in the interior of the stable tube and the
flow will proceed along the stable tube towards the neighborhood of the
critical point $E$, from where it will re-enter the unstable tube and proceeds
towards the FRW singularity and by a new intersection a part of these orbits
will again enter the stable tube and proceed back towards the neighborhood of
the critical point $E$, and so on, producing an infinite recurrence of the
motion for a class of orbits, which will be periodic orbits of very long
periods and bounded oscillatory orbits. Another part of the orbits inside the
unstable tube will flow along the exterior of the stable tube towards the
neighborhood of the critical point $E$, and afterwards will escape towards the
DeSitter attractor at infinity along the exterior of the unstable tube of the
second pair (cf. Fig. 2). This for one single intersection. The same pattern
will be reproduced for each of the infinite intersections of the unstable tube
with the stable one. In this process, the surface of the tubes, which is in
fact a boundary for distinct types of flows, will become more and more
stretched and folded, resulting in an intricate structure which will have some
important physical consequences for the long time behaviour of the orbits.
Indeed,  basins of initial conditions for initially expanding universes have a
fractal boundary associated with recollapse or escape into the DeSitter
configuration, as we show soon.

We must comment that the chaotic aspect of the dynamics described here depends
crucially on the presence of the cosmological constant that engenders the
structure of a saddle-center-center critical point in phase space together with
the homoclinic four dimensional tubes, the transversal crossings of which (in a
neighborhood of the FRW singularity) characterize chaos definitely and
unambiguously in the model. If we restrict ourselves to the energy surface
$E_0=0$ we have the \textit{Mixmaster universe with a cosmological constant}
which presents the same structure of chaos. For the limiting case ($E_0=0,
\Lambda=0$), the absence of the critical point appears to eliminate the set of
homoclinic orbits and periodic orbits of arbitrarily large periods, whose
recurrence engenders chaos in the models. In this instance, our approach does
not allow to invariantly characterize chaos in the Mixmaster model.

Our system is open (noncompact) with two definite asymptotic exits, namely,
collapse and escape to the DeSitter configuration. The code escape/collapse
defines basin boundaries in the initial conditions set; these boundaries are
associated precisely to the surface of the homoclinic tubes in the set, and
correspond to the initial conditions for the homoclinic intersection manifold
contained simultaneously in the stable and in the unstable 4-dim tubes.
Together with the countable set of periodic orbits of arbitrarily large periods
that exist in the neighborhood of each homoclinic \textit{orbit} and that have
the homoclinic \textit{orbit} as an accumulation set
\cite{ozorio}\cite{ozorio1}, they constitute the set of orbits which neither
escape or collapse. This set of bounded orbits that are in the boundary of
collapse and escape constitutes , in this way, the \textit{strange repellor}.
The fractal dimension of the basin boundary sets are calculated below by a
box-counting method.

\begin{figure}[htb]
\hspace{-1cm}
\rotatebox{270}{\includegraphics*[scale=0.3]{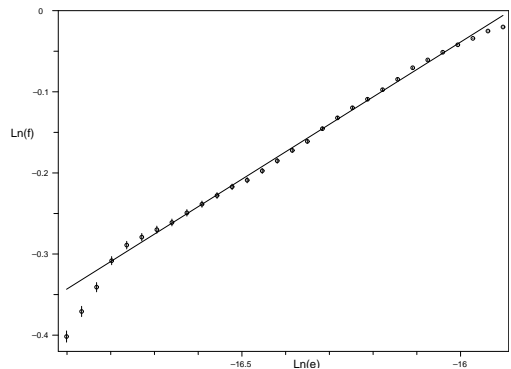}}
\caption{Plot of the scaling law $f\propto \epsilon^{\alpha}$ in log-log scale, where
$\epsilon$ is the uncertainty radius about 5,000 initial conditions taken
inside the set $\cal D$ constructed about the point
$(x=0.4,y=1,z=1,p_x=2.036467455281656,p_y=0,p_z=0)$ on the separatrix of the
invariant plane $(x,p_x)$. $f$ is the fraction of $\cal D$ of uncertain initial
conditions with respect to the uncertainty code collapse/escape. The set $\cal
D$ is a hypercube of edges $10^{-4}$ about the mentioned point, in the energy
surface $E_0=0.9999999683772234$. The best fit of the small linear region
renders $(a)$ $\alpha\approx 0.321$.}
\end{figure}

In other words, our system has the characteristic of a chaotic scattering
system, having the saddle-center-center as a chaotic scatterer, with two
absolute outcomes consisting of (i) escape into inflation (the DeSitter
attractor) or (ii) recollapse to the singularity, for initially expanding
universes. The DeSitter attractor defines a way out from the initial
singularity to the inflationary phase but this exit to inflation is completely
chaotic, namely, small fluctuations in initial conditions may cause the
universe to change its asymptotic behaviour from recollapse to escape into
inflation and vice-versa. This is due to the presence of chaotic sets in the
phase space of the system, as a consequence of the structure of homoclinic
tubes and their transversal crossings, which provides an invariant
characterization of chaos in the model. One of the questions to be examined
here is the characterization of sets of initial conditions for which the
DeSitter attractor is attained and establish the character of a chaotic
scattering system. To see this chaotic exit to inflation, let us consider
initial conditions sets for initially expanding universes in the energy surface
$E_0$ such that the orbits visit a linear neighborhood of the
saddle-center-center critical point $E$. In this neighborhood, the Hamiltonian
is separable according to (\ref{eq12}), namely,
$H=(E_{crit}-E_0)+E_{hyp}-E_{rot_{1}}-E_{rot_{2}}=0$. The \textit{partition} of the
energy $(E_{crit}-E_0)$ into the energies $E_{hyp}$ and $E_{rot_{1}}+E_{rot_{2}}$ of
motion about the critical point will determine the outcome of the oscillatory
regime about $E$ into collapse or escape to inflation (DeSitter attractor)
whether $E_{hyp}<0$ or $E_{hyp}>0$, respectively. However the nonintegrability
of the system, with the consequent homoclinic crossing of the tubes, will cause
this \textit{partition} to be chaotic in general, namely, given an arbitrary
initial condition of energy $E_0$ we are no longer able to foretell in which of
regions $I$ or $II$ about the saddle-center the orbit will land when it
approaches $E$. Since $E_{hyp}=0$ is a limiting case, for initially expanding
universes, the motion inside the tubes corresponds to orbits that will
recollapse after reaching the neighborhood of $E$, while the motion outside the
tubes corresponds to orbits that will escape into the inflationary phase
(towards the DeSitter attractor at infinity). In other words, the intrincate
crossing and merging of the tubes produces chaotic sets in the phase space of
the model, in particular establishing fractal basin boundaries associated with
the chaotic exit to inflation. To illustrate this, we finally present a
numerical experiment where we obtain a measure of the fractal dimension of
basin boundaries in sets of initial conditions by using the box-counting method
with the uncertainty code collapse/escape into inflation. In this numerical
experiment we evaluate the fractal dimension of portions $\cal{D}$ of phase
space, about a point of the separatrix $S$ on the invariant plane. The
box-counting method, due to Ott and collaborators \cite{ott}, consists of
determining the uncertainty exponent $\alpha$ appearing in the scaling law
$f\propto \epsilon^{\alpha}$, where $\epsilon$ is the uncertainty radius about
5,000 initial conditions taken inside the sets $\cal{D}$, and $f$ is the
fraction of $\cal{D}$ of uncertain initial conditions with the uncertainty code
collapse/escape into inflation. The uncertainty exponent is related to the
fractal dimension $d$ of the fractal basin boundary by $d=N-\alpha$, where
$N=5$ is the phase space energy surface dimensionality. In our numerical
experiment we selected a set $\cal{D}$ of homogeneously and randomly
distributed initial conditions about the point
$(x=0.4,y=1,z=1,p_x=2.036467455281656,p_y=0,p_z=0)$ on the separatrix $S$, such
that the orbits visit a neighborhood of the saddle-center-center $E$ before
collapse or escape into inflation. In Fig. 5 we depict the scaling law
$f\propto \epsilon^{\alpha}$ in which the small linear portion lies between the
saturation region for large $\epsilon$ and noise for very small $\epsilon$. The
following value for the fractal dimension of the set $\cal{D}$ is found,
$d\approx 4.627$, confirming the chaotic nature of these sets.

\acknowledgments

The authors acknowledge CNPQ/Brazil for partial financial support.

\end{document}